\documentclass[a4paper,12pt]{article}
\usepackage{epsfig}
\def\be{\begin{equation}}
\def\ee{\end{equation}}
\def\bc{\begin{center}}
\def\ec{\end{center}}
\def\bea{\begin{eqnarray}}
\def\eea{\end{eqnarray}}

\def\ov{\overline}
\def\cO{{\cal O}}
\def\cL{{\cal L}}
\def\gold{{\widetilde{G}}}
\def\simlt{\mathrel{\lower2.5pt\vbox{\lineskip=0pt\baselineskip=0pt
              \hbox{$<$}\hbox{$\sim$}}}}
\def\simgt{\mathrel{\lower2.5pt\vbox{\lineskip=0pt\baselineskip=0pt
              \hbox{$>$}\hbox{$\sim$}}}}
\renewcommand{\theequation}{\thesection.\arabic{equation}}
\title{
\vspace*{-0.8cm}
\begin{flushright}
\normalsize{
CERN--PH--TH/2004--195
\\
ROMA-1390/04 
}
\\
\end{flushright}
\vspace{1cm}
\Large\textbf{Phenomenology of a leptonic goldstino \\
and invisible Higgs boson decays}
\author{\large
{\bf I.~Antoniadis~$^1$\footnote{On leave of absence from CPHT,
Ecole Polytechnique, UMR du CNRS 7644.}~,
M.~Tuckmantel~$^{1,2}$,
F.~Zwirner~$^{1,3}$}
\\ \\
\emph{$^1$Department of Physics, CERN - Theory Division}\\
\emph{CH--1211 Geneva 23, Switzerland}\\
\emph{$^2$Institut f\"ur Theoretische Physik, ETH H\"onggerberg}\\
\emph{CH--8093\, Z\"urich, Switzerland}\\
\emph{$^3$Dipartimento di Fisica, Univ. di Roma La Sapienza, and}\\
\emph{INFN, Sez. di Roma, P.le A.~Moro 2, I--00185 Rome, Italy}}}
\date{}

\catcode`@=11
\def\marginnote#1{}
\newcount\hour
\newcount\minute
\newtoks\amorpm
\hour=\time\divide\hour by60
\minute=\time{\multiply\hour by60 \global\advance\minute by-\hour}
\edef\standardtime{{\ifnum\hour<12 \global\amorpm={am}%
        \else\global\amorpm={pm}\advance\hour by-12 \fi
        \ifnum\hour=0 \hour=12 \fi
        \number\hour:\ifnum\minute<10 0\fi\number\minute\the\amorpm}}
\edef\militarytime{\number\hour:\ifnum\minute<10 0\fi\number\minute}
\def\draftlabel#1{{\@bsphack\if@filesw {\let\thepage\relax
   \xdef\@gtempa{\write\@auxout{\string
      \newlabel{#1}{{\@currentlabel}{\thepage}}}}}\@gtempa
   \if@nobreak \ifvmode\nobreak\fi\fi\fi\@esphack}
        \gdef\@eqnlabel{#1}}
\def\@eqnlabel{}
\def\@vacuum{}
\def\draftmarginnote#1{\marginpar{\raggedright\scriptsize\tt#1}}
\def\draft{\oddsidemargin 0.0truein
        \def\@oddfoot{\sl preliminary draft \hfil
        \rm\thepage\hfil\sl\today\quad\militarytime}
        \let\@evenfoot\@oddfoot \overfullrule 3pt
        \let\label=\draftlabel
        \let\marginnote=\draftmarginnote
   \def\@eqnnum{(\theequation)\rlap{\kern\marginparsep\tt\@eqnlabel}%
\global\let\@eqnlabel\@vacuum}  }
\catcode`@=12
\begin{document}
\maketitle
\thispagestyle{empty}
\vspace*{.5cm}
\begin{abstract}
Non-linearly realized supersymmetry, combined with the Standard Model
field content and $SU(3) \times SU(2) \times U(1)$ gauge invariance,
permits local dimension-six operators involving a goldstino, a lepton
doublet and a Higgs doublet. These interactions preserve total lepton
number if the left-handed goldstino transforms as an antilepton. We
discuss the resulting phenomenology, in the simple limit where the new
couplings involve only one lepton family, thus conserving also lepton
flavour. Both the $Z$ boson and the Higgs boson can decay into a
neutrino and a goldstino: the present limits from the invisible $Z$
width and from other observables leave room for the striking
possibility of a Higgs boson decaying dominantly, or at least with a
sizable branching ratio, via such an invisible mode. We finally
comment on the perspectives at hadron and lepton colliders, and on
possible extensions of our analysis.
\end{abstract}
%
\date
\newpage
\section{Introduction} 
\label{introduction}
\subsection{General theoretical framework} 
The first attempt to introduce supersymmetry in a particle physics
model dates back to the seminal paper by Volkov and Akulov
\cite{volaku}: simple four-dimensional supersymmetry was non-linearly
realized, the goldstino was identified with the (electron) neutrino,
and a universal dimension-eight coupling of goldstino bilinears to the
matter energy-momentum tensor was introduced, with strength fixed by
the goldstino decay constant. Soon after, it was realized
\cite{bardwf} that interpreting the goldstino as one of the Standard
Model (SM) neutrinos was not allowed by the low-energy theorems of
supersymmetry, which prescribe a much softer infrared behaviour of the
goldstino amplitudes with respect to the neutrino ones. Many years
later, it was also noticed that, already at the lowest order in the
derivative expansion, the bilinear goldstino couplings to the SM
fields are more general than the universal coupling to the
energy-momentum tensor \cite{bfzfer, cllw, lutpon}, and explicit
examples were produced \cite{abl} in superstring models with 
D-branes.

Recently, another unexpected result was found for the single-goldstino
couplings \cite{anttuc}, again in the context of superstring models
with D-branes. In the linear realization of spontaneously broken $N=1$
supersymmetry, a single goldstino couples universally to the
supersymmetry current, in a way prescribed by the supersymmetry
algebra \cite{fayet}. When moving to the non-linear realization
\cite{volaku, nonlin}, for example by integrating out all the heavy
superpartners of the goldstino and of the remaining light fields, all
these couplings are expected to disappear, leaving only interactions
with an even number of goldstinos. However, as recently found in
\cite{anttuc}, this is not the most general possibility: for a generic
field content, there are two types of operators of dimension six, with
undetermined coefficients, that contain a single goldstino coupled to
a matter fermion and a gauge or scalar field. 

Further restrictions are obtained by assuming the SM gauge group and
field content, in addition to a gauge-singlet goldstino. At the lowest
order in the goldstino decay constant $\kappa$, which sets the scale
of supersymmetry breaking, the only dimension-six operator that can
couple one goldstino to SM fields is:
\be
\cO = \kappa \ \sum_{a=e,\mu,\tau} c_a \ \cO_a + h.c. \, ,
\qquad
\cO_a = \epsilon_{ij} \ l_a^{\, i} \ (\partial^\mu 
\widetilde{G}) \ (D_\mu \phi)^{\, j} \, .
\label{operator}
\ee
In eq.~(\ref{operator}), $\kappa$ is a real coefficient of dimension
$(mass)^{-2}$, whose normalization is determined, for example, by the
inhomogeneous term in the goldstino transformation law, $\delta_{\xi}
\widetilde{G}_{\alpha} = (1 / \kappa) \, \xi_{\alpha} + \ldots$ The
$c_a$ are generically complex dimensionless parameters, where the
index $a=e,\mu,\tau$ denotes the three different lepton families. The
fields appearing in the operators $\cO_a$ are the goldstino $\gold
\sim (1,1,0)$, the lepton doublets $l_a^{\, i} \equiv (\nu_a,e_a)^T
\sim (1,2,-1/2)$ and the SM Higgs doublet $\phi^{\, i} \equiv
(\varphi^+,\varphi^0)^T \sim (1,2,+1/2)$, where the numbers in
brackets denote the transformation properties with respect to $SU(3)_C
\times SU(2)_L \times U(1)_Y$, and we work with two-component spinors,
in the notation of ref.~\cite{wesbag}. For convenience, it is not
restrictive to work in a field basis where all kinetic terms are
canonically normalized, and the charged leptons $e_a$ are in a mass
eigenstate basis. The symbol $\epsilon_{ij}$ is the $SU(2)_L$
antisymmetric tensor, normalized according to $\epsilon_{1 2} =
1$. Finally, $D_\mu$ is the SM gauge-covariant derivative. In this
paper, we restrict ourselves to the couplings of the goldstino with
the minimal SM, including the Higgs field but neither right-handed
neutrinos nor dimension-five operators inducing Majorana masses for
the left-handed neutrinos: we will briefly comment on the consequences
of relaxing such simplifying assumption in the final section.

\subsection{String models with D-branes}

We have already mentioned that operators of the form (\ref{operator})
were recently found \cite{anttuc} in superstring models with D-branes.
Although our main considerations below will have general validity,
irrespectively of the microscopic origin of the operator
(\ref{operator}), we describe here for illustration the main
properties of the goldstino in such a D-brane string framework.
Readers interested only in the phenomenological description at the
effective field theory level can skip this subsection and go directly
to subsection~1.3.

These higher-dimensional string constructions involve configurations
of intersecting branes, combined eventually with orientifolds (for
recent reviews on semirealistic D-brane models, and references to the
original literature, see e.g. \cite{bramod}). In this context, there
are (at least) two bulk supersymmetries that come into play. One
(half) is spontaneously broken by the very existence of the branes,
while the other one (other half) is broken by the fact that the branes
are at angles, or by the simultaneous presence of orientifolds.  The
goldstino appearing in the four-dimensional effective theory is
associated with the former, not with the latter \cite{abl,anttuc}.
Since the corresponding supersymmetry is broken by the presence of the
branes, it can only be realized non linearly, and brane fields have no
superpartners.  For the simple case of two stacks of D-branes at
angles, the goldstino decay constant $\kappa$ is given by the total
brane tension on their intersection. Moreover, the coefficients $c_a$
turn out to be real, universal model-independent constants: $c_a=0$
(if the lepton and the Higgs doublet come from different intersections)
or $|c_a| = 2$. The possibility of having observable effects at the
presently accessible energies is then related with the possibility of
having the string scale close to the weak scale \cite{lowss}.

More precisely, for two stacks of $N_1$ coincident and $N_2$
coincident D-branes, intersecting in a $3+1$ dimensional volume, we
have~\cite{anttuc}:
\begin{equation}
{1\over 2 \ \kappa^2} = N_1 \, T_1 + N_2 \, T_2 \, ;
\qquad
T_i ={M_s^4 \over 4 \pi^2 g_i^2} \, ,
\quad
(i=1,2) \, ,
\label{decayconst}
\end{equation}
where $M_s \equiv (\alpha^\prime)^{-1/2}$ is the string scale, $T_1$
and $T_2$ are the effective tensions at the intersection, and $g_i$
($i=1,2$) is the effective four-dimensional gauge coupling on the
$i$-th brane stack. To get an estimate of the ratio between the string
scale and the supersymmetry-breaking scale, $\sqrt{F} \equiv
[1/(2\kappa^2)]^{1/4}$, we assume that the Higgs and lepton doublets
come from the intersection of an abelian brane ($N_1=1$) with two
coincident branes ($N_2=2$), describing the $SU(2)_L$ factor of the SM
gauge group. To remove the ambiguity on the $U(1)$ coupling $g_1$,
related to the hypercharge embedding, we can use the representative
GUT values $g_1 \simeq g_2 \simeq 1/ \sqrt{2}$: in such a case $M_s
\simeq 1.6 \ \sqrt{F}$. For $g_1 \gg g_2$ we could reach $M_s \simeq
1.8 \ \sqrt{F}$. Too small values of $g_1$ would be phenomenologically
incompatible with values of $\sqrt{F}$ close to the weak scale.

In our analysis, we work in the limit of global supersymmetry and
neglect gravitational effects. When allowed by the above string
constructions, this amounts to taking the decompactification limit in
a direction transverse to the brane configuration, which suppresses
the four-dimensional gravitational interactions. At finite volume, the
non-linearly realized supersymmetry is in general broken when
including effects coming from different locations of the bulk, thus
our brane-localized goldstino is expected to acquire a mass suppressed
by a volume factor. In particular, our putative goldstino should mix
with the internal components of the higher-dimensional gravitino, but
the resulting mass terms are expected to be small enough to be
irrelevant for the purposes of the present paper.

\subsection{Effective theory setup}

In summary, we are considering an effective theory where the $SU(3)_C
\times SU(2)_L \times U(1)_Y$ gauge invariance is linearly realized,
whilst global $N=1$ supersymmetry is non-linearly realized. Its
content amounts to the fields of the minimal, non-supersymmetric SM,
plus a gauge-singlet goldstino $\gold$. Its Lagrangian is given by
\be
\cL = \cL_{SM} - {i \over 2} [ \gold \
\sigma^\mu \ \partial_\mu \ \ov{\gold} - (\partial_\mu \ \gold ) 
\ \sigma^\mu \ \ov{\gold} ] + \cO + \ldots \, , 
\label{lag}
\ee
where $\cL_{SM}$ is the renormalizable SM Lagrangian (including a
possible cosmological constant term), $\cO$ is the one-goldstino
operator in (\ref{operator}), and the dots indicate additional terms,
containing at least two goldstino fields and of higher order in
$\kappa$.

It is important to notice that the effective theory defined by
(\ref{operator}) and (\ref{lag}) conserves the total lepton number
$L$, as long as we assign to the left-handed goldstino $\gold$ a total
lepton number $L(\gold)=-1$. In a sense, this model partially
implements the original proposal of \cite{volaku}: the goldstino is
not identified with a SM neutrino, but can nevertheless be regarded as
a neutral gauge-singlet (anti-)lepton. The analogy between lepton
number and a continuous $R$-symmetry, already noticed in \cite{fayet},
becomes now an identification: the conserved lepton number of our
effective theory is associated with a diagonal subgroup of
$U(1)_{\widehat{L}} \times U(1)_R$, where $U(1)_{\widehat{L}}$
corresponds to the SM lepton number, and acts on the leptons but not
on the goldstino, whilst $U(1)_R$ is the $R$-symmetry, acting on the
goldstino but not on the leptons.

The goal of the present paper is to discuss the phenomenological
implications of the new operator in eq.~(\ref{operator}), under the
simplifying assumption that only one of the coefficients $c_a$ is
non-vanishing:
\be
\label{assum}
c_{\, \widehat{a}} = c \ne 0 \quad 
{\rm for} \; {\rm one} \; \widehat{a} \, ,
\qquad \quad
c_a=0 \quad {\rm for} \; a \ne \widehat{a} \, .
\ee
This assumption is very close in spirit to a similar one, made in most
phenomenological studies of explicit $R$-parity violation in the Minimal
Supersymmetric Standard Model (MSSM). The fact that in the SM the
Higgs boson couples to fermions proportionally to their masses, and
the geometrical setup of D-brane models, suggest that the
non-vanishing coupling in (\ref{assum}) is more likely associated with
the heavier lepton generations. However, in most of the following
considerations we will treat all three possible choices of
$\widehat{a}$ in (\ref{assum}) on equal footing.

Under the assumption (\ref{assum}), the operator $\cO$ conserves not
only the total lepton number $L$, but also the partial lepton numbers
$L_a$, as long as we assign to the goldstino $\gold$ partial lepton
numbers $L_{\, \widehat{a}} (\gold) = - 1$ and $L_a(\gold) = 0$ for $a
\ne \widehat{a}$. This will allow us to discuss, in the central
section of the paper, only processes conserving both total and partial
lepton numbers, in the limit of vanishing neutrino masses. We will
first review the phenomenological constraints on the new goldstino
couplings coming from known physics: under our assumptions, the most
stringent ones come from the LEP bounds on the invisible $Z$ width.
Other constraints either are weaker or have a more ambiguous
interpretation within the effective theory. We will then discuss a
striking phenomenological implication of the new couplings: the
possibility of having, as non-negligible or even dominant decay mode
for the Higgs boson, the invisible channel consisting of a neutrino
and a goldstino (or the conjugate channel). In the final section we
will present our conclusions, and comment on the experimental
perspectives at high-energy colliders and on possible generalizations
of the present analysis.

\section{Phenomenology}
\setcounter{equation}{0}

We now discuss the phenomenology of the operator $\cO$ of
eq.~(\ref{operator}), under the assumption that, in a field basis
where the charged leptons are mass eigenstates, only one of the
coefficients $c_a$ is non-negligible. We can then omit the generation
index $a$ and recall that
\be
D_\mu \phi = \partial_\mu \phi - i \ \left( g \ {\sigma^I \over 2} \ 
A_\mu^I + g' \ {1 \over 2} \ B_\mu \right) \ \phi \, ,
\ee
where $\sigma^I$ ($I=1,2,3$) are the Pauli matrices, $g$ and $g'$ are 
the $SU(2)_L$ and $U(1)_Y$ gauge coupling constants, respectively, 
and $A_\mu^I$ ($I=1,2,3$) and $B_\mu$ the corresponding gauge bosons.
We can also move to the unitary gauge, where
\be
\phi = \left( \begin{array}{c}
\varphi^+ \\ \varphi^0 \end{array} \right) 
= {1 \over \sqrt{2}} \ \left( \begin{array}{c} 0 \\ 
v + H \end{array} \right) \, ,
\ee
$H$ is the canonically normalized SM Higgs boson, and the vacuum
expectation value $v \equiv \sqrt{2} \, \langle \varphi^0 \rangle$ is
related with the Fermi coupling and the gauge boson masses by
\be
{G_F \over \sqrt{2}} = {1 \over 2 \ v^2} =
{g^2 \over 8 \ m_W^2} = {g^2 + g^{\prime \, 2} \over
8 \ m_Z^2} \, .
\ee
Recalling that
\be
W_\mu^\pm = {A_\mu^1 \mp i \ A_\mu^2 \over \sqrt{2}} \, ,
\qquad 
Z_\mu = {g' \ B_\mu - g \ A_\mu^3 \over \sqrt{g^2 + 
g^{\prime \, 2} }} \, ,
\ee
we can then write
\bea
\cO & = & \left[
{\kappa \ c \over \sqrt{2}} \ \nu \
(\partial^\mu \gold) \ 
 (\partial_\mu H) + h.c. \right]
\label{uno} \\
& - & \left[
{i \ \kappa \ c \over \sqrt{2}} \ m_Z \ \nu 
\ (\partial^\mu \gold) \ Z_\mu + h.c. \right]
\label{due} \\
& + & \left[
i \ \kappa \ c \ \ m_W \ e  \ (\partial^\mu \gold) 
\ W_\mu^+ + h.c. \right] 
\label{tre} \\
& - & \left[
{i \ \kappa \ c \ \sqrt{g^2 + g^{\prime \, 2}}
\over 2 \ \sqrt{2}} \ \nu \ 
(\partial^\mu \gold) \ Z_\mu \ H + h.c. \right]
\label{quattro} \\
& + & \left[
{i \ \kappa \ c \ g \over 2} \ e \ 
(\partial^\mu \gold) \ W_\mu^+ \ H + h.c. \right]
\label{cinque} \, .
\eea

Before examining the phenomenological consequences of the interactions
(\ref{uno})-(\ref{cinque}), it is useful to recall the general
experimental limits on the supersymmetry breaking scale. Under the
assumption that all superpartners of the goldstino and of the SM
particles are sufficiently heavy to have escaped detection, which
corresponds to the non-linear realization, the most stringent bounds
come from the processes $e^+ e^- \rightarrow \gold \ \ov{\gold} \
\gamma$ at LEP2 \cite{nrw, lepth, lepbound}, and give $\sqrt{F} >
238$~GeV ($95 \, \%$ c.l.). The study of the processes $p \ov{p}
\rightarrow \gold \ \ov{\gold} \ \gamma$, $\gold \ \ov{\gold} \ jet$
at the Tevatron collider \cite{nrw, tevth} has led so far to a
published bound \cite{tevbound} $\sqrt{F} > 221$~GeV ($95 \, \%$
c.l.), lower than the LEP2 bound. Finally, indirect bounds from the
muon anomalous magnetic moment \cite{gm2bound}, from flavour physics
\cite{flabound} and from astrophysics and cosmology \cite{astbound}
are less stringent than the present collider bounds~\footnote{At
least this is true in the absence of (light) superpartners of the SM
particles and of the goldstino, as is the case in the present
theoretical framework.}.

Inspection of eqs.~(\ref{uno})-(\ref{cinque}) shows that the new local
interactions in (\ref{due}) and (\ref{tre}) can describe non-standard
contributions to measured processes, where an antineutrino is replaced
by a goldstino (a neutrino by an antigoldstino) in the final state.
These contributions occur already at order $\kappa$ in the amplitude,
thus $\kappa^2$ in the cross-section or in the decay rate. In
addition, the new local interactions in (\ref{uno}), (\ref{quattro})
and (\ref{cinque}) can describe, at the same order in $\kappa$, decays
of the SM Higgs boson into final states containing leptons and
goldstinos.

\subsection{Constraints from known physics}

Leaving the study of Higgs decays for the next subsection, we now
discuss the phenomenological constraints coming from known
physics. Whenever possible, it will be convenient to express these
constraints in terms of the auxiliary mass parameter
\be
\label{mdef}
M \equiv 
{1 \over \sqrt{|\kappa \, c |}}
= 2^{1/4} \, \sqrt{F \over  | c |} \, . 
\ee

The interaction in (\ref{due}) describes the decays $Z \rightarrow \nu
\, \widetilde{G}$ and $Z \rightarrow \ov{\nu} \, \ov{\widetilde{G}}$.
Their signal would be an additional contribution to the $Z$ partial
width into invisible products, besides the one associated with the
SM neutrinos:
\be
\Delta \Gamma_{inv} (Z) =
\Gamma ( Z \rightarrow \nu \, \widetilde{G} ) + 
\Gamma ( Z \rightarrow \ov{\nu} \, \ov{\widetilde{G}} ) =
{m_Z^5 \over 192 \ \pi \ M^4} \, .
\label{zdecay}
\ee
The present upper bound on exotic contributions to the invisible $Z$
width is \cite{lepewwg}:
\be
\label{gzinv}
\Delta \Gamma_{inv} (Z)  < 2.0 \ {\rm MeV} \, ,
\qquad
(95 \, \% \ {\rm c.l.}) \, .
\ee
Plugging this into eq.~(\ref{zdecay}), we obtain:
\be
M > 270 \ {\rm GeV} \, . 
\label{zinvbound}
\ee
For specific values of $c$, we can extract a bound on the
supersymmetry breaking scale and compare it with the collider
bounds. For example, taking $|c| = 2$ as suggested by the results of
\cite{anttuc}, we get $\sqrt{F} > 320 \ {\rm GeV}$, slightly stronger
than the present collider bounds~\footnote{Collider bounds on
$\sqrt{F}$ in the non-linear realization have a very mild dependence
on the free parameters of the ${\cal O}(\kappa^2)$ four-fermion
couplings between two SM fermions and two goldstinos \cite{bfzfer,
abl, lepth, anttuc}. This dependence is negligible for the present
analysis.}.

The interaction in (\ref{tre}) describes the decays $W^+ \rightarrow
\ell^+ \, \ov{\widetilde{G}}$ and $W^- \rightarrow \ell^- \,
\widetilde{G}$, where $\ell=e,\mu,\tau$ according to the choice of
$\widehat{a}$ in (\ref{assum}), with partial widths (neglecting the
charged lepton mass):
\be
\Gamma ( W^+ \rightarrow \ell^+ \, \ov{\widetilde{G}} ) = 
\Gamma ( W^- \rightarrow \ell^- \, \widetilde{G} ) =
{m_W^5 \over 192 \ \pi \ M^4} \, .
\label{wdecay}
\ee
The constraints from $W$ decays are similar to those from $Z$ decays,
but weaker. Under the assumption (\ref{assum}), exotic $W$ decays of
the type (\ref{wdecay}) could produce violations of lepton
universality, via their additional contributions to one of the
leptonic widths. Taking $M$ at its lower bound (\ref{zinvbound}), we
would find $\Delta \Gamma_\ell (W) \simeq 1 \ {\rm MeV}$,
corresponding to $\Delta BR_\ell (W) \simeq 4.5 \times 10^{-4}$, still
below the present precision of the LEP2 and Tevatron experiments
\cite{lepewwg}.

The non-renormalizable nature of our new $d=6$ operator suggests that
its effects have a strong, power-like suppression when the typical
energy scale of the processes under consideration is much smaller than
$M$. However, we should check that precisely measured low-energy
processes, such as $\mu$ and $\tau$ decays, cannot give constraints
stronger than (\ref{zinvbound}). To give an idea of the sensitivity in
$\mu$ and $\tau$ decays, we recall the present experimental precision
\cite{pdg} on the respective rates (or, equivalently, on the
lifetimes):
\be
\left| {\Delta \Gamma_\mu \over \Gamma_\mu} \right|
\sim 2 \times 10^{-5} \, ,
\qquad
\left| {\Delta \Gamma_\tau \over \Gamma_\tau} \right|
\sim { 1 \over 300} \, .
\label{mutauexp}
\ee
Exotic contributions to $\mu$ and $\tau$ decays can be originated by
the new charged current interactions of eq.~(\ref{tre}). The new
Feynman diagrams involve a goldstino at the place of an antineutrino
(or an antigoldstino at the place of a neutrino) on an external line,
thus there is no interference with the SM ones. The additional
contributions to the decay rates scale then at most as
\be
\left| {\Delta \Gamma_\ell \over \Gamma_\ell} \right|
\simlt  {m_\ell^2 \ m_W^2 \over g^2 \ M^4} \, ,
\qquad
(\ell=\mu \, , \tau) \, .
\label{mutauth}
\ee
Taking $M$ at its lower bound (\ref{zinvbound}), we would find $\Delta
\Gamma_\mu / \Gamma_\mu \simlt 3 \times 10^{-8}$ and $\Delta
\Gamma_\tau / \Gamma_\tau \simlt 10^{-5}$, orders of magnitude below
the present experimental sensitivity. Hadron decays with
(semi-)leptonic final states, neutrino-electron scattering and
neutrino-hadron deep inelastic scattering are also expected, on the
basis of similar reasonings, to provide weaker constraints than
(\ref{zinvbound}).

Another process sensitive to the new couplings in (\ref{due}) and
(\ref{tre}) is $e^+ e^- \to \gamma + {\rm nothing}$ at LEP2. With two
goldstinos in the final state, this process has an amplitude ${\cal
O}(\kappa^2)$, and is used \cite{nrw, lepth, lepbound} to establish
the model-independent lower bound on the supersymmetry-breaking scale.
With one neutrino and one goldstino (one antineutrino and one
antigoldstino) in the final state, the amplitude for this process
occurs at ${\cal O}(\kappa)$, and may receive two contributions: the
one from (\ref{due}), corresponding to Z exchange in the s-channel, is
always present under our assumptions; the one from (\ref{tre}),
corresponding to W exchange in the t-channel, is present if and only
if the new coupling involves the first generation ($\widehat{a} =
e$). To be conservative, we can estimate the bound on $M$, defined in
(\ref{mdef}), by assuming that (\ref{due}) does contribute, but
(\ref{tre}) does not. According to the formalism of \cite{radiator},
we can approximate the differential cross-section for the processes
$e^+ e^- \to \gamma \, \nu \, \widetilde{G}$ and $e^+ e^- \to \gamma
\, \ov{\nu} \, \ov{\widetilde{G}}$, dominated by the soft and
collinear part of the photon spectrum ($x_\gamma \ll 1$ and/or $\sin^2
\, \theta_\gamma \ll 1$), by:
\be
{d \sigma \over d x_\gamma \, d \cos \theta_\gamma}
\simeq
\sigma_0(\widehat{s}) \, {\alpha \over \pi} \, 
{1 + (1 - x_\gamma)^2 \over x_\gamma \, \sin^2 \theta_\gamma}
\, ,
\label{dsph}
\ee
where $x_\gamma$ is the fraction of the beam energy carried by the
photon, $\theta_\gamma$ is the scattering angle of the photon with
respect to the direction of the incoming electron in the
centre-of-mass frame, $\widehat{s} = (1 - x_\gamma) \, s$, $\sqrt{s}$
is the energy in the centre-of-mass frame, and
\be
\sigma_0 \equiv \sigma \left( e^+ e^- \to 
\nu \, \widetilde{G} \right)
+
\sigma \left( e^+ e^- \to \ov{\nu} 
\, \ov{\widetilde{G}} \right) \, . 
\label{sigma0}
\ee
In the conservative case where only the coupling (\ref{due}) is
present, we can write
\be
\label{sigma0th}
\sigma_0 ( \widehat{s} ) = {12 \, \pi \, \widehat{s}^2 \over
m_Z^4} \, {\Gamma_e \ \Delta \Gamma_{inv}(Z) \over ( \widehat{s}
- m_Z^2 )^2 + \Gamma_Z^2 \, m_Z^2 } \, ,
\ee
where $\Gamma_e \equiv \Gamma ( Z \to e^+ e^-) \simeq 84 \; {\rm
MeV}$. We can approximate the present LEP2 sensitivity \cite{lepbound}
by requiring that, for $\sqrt{s} = 207 \; {\rm GeV}$, $x_\gamma >
0.05$ and $| \cos \, \theta_\gamma | < 0.95$, it is $\sigma ( e^+ e^-
\to \gamma \, \nu \, \widetilde{G}) + \sigma ( e^+ e^- \to \gamma \,
\ov{\nu} \, \ov{\widetilde{G}}) < 0.1 \; {\rm pb}$. Plugging $M > 270
\; {\rm GeV}$ from eq.~(\ref{zinvbound}) into the expression
(\ref{zdecay}) for $\Delta \Gamma_{inv}(Z)$, and making use of
eqs.~(\ref{dsph})--(\ref{sigma0th}), we find
\be
\sigma ( e^+ e^- \to  \gamma \, \nu \, \widetilde{G})
+
\sigma ( e^+ e^- \to \gamma \, \ov{\nu} \, 
\ov{\widetilde{G}}) < 0.011 \; {\rm pb} \, ,
\ee
roughly one order of magnitude below the LEP2 sensitivity. The
possible inclusion of the $W$-exchange diagram, associated with the
interaction (\ref{tre}), should not modify significantly the above
conclusion.

We can try to extract additional constraints on $M$ by considering
high-energy SM processes where virtual goldstinos are exchanged on the
internal lines. To extract reliable constraints, however, we should
be sure that there are no additional local operators contributing to
the total amplitude for the same process at ${\cal O}(\kappa^2)$. In
the absence of a microscopic theory, such bounds should be regarded
only as order of magnitude estimates.

An example is $e^+ e^- \to W^+ W^-$ at LEP2. At the classical level,
and in the limit where the electron mass is neglected, the SM
amplitude receives contributions from three Feynman diagrams: one with
the $t$-channel exchange of the electron neutrino, and two with the
$s$-channel exchange of the photon and the $Z$. If the new interaction
(\ref{tre}) does not involve the first lepton family ($ \widehat{a} =
\mu, \, \tau$), there is of course no exotic contribution, thus no
constraint. Otherwise ($\widehat{a}=e$), the new interaction generates
an ${\cal O}(\kappa^2)$ amplitude, corresponding to a Feynman diagram
with goldstino exchange in the $t$-channel, that interferes with the
SM diagrams. To estimate the order of magnitude bound from $W$-pair
production at LEP2, we decompose the total cross-section $\sigma_{WW}$
for the (on-shell) process as:
\be
\label{sigmaww}
\sigma_{WW} = \sigma_{SM} + 
\sigma_{\kappa^2} + \sigma_{\kappa^4} \, .
\ee
Here $\sigma_{WW}$ is the SM cross-section, $\sigma_{\kappa^2}$ is the
${\cal O} (\kappa^2)$ interference contribution, and $\sigma_{
\kappa^4}$ is the ${\cal O} (\kappa^4)$ contribution from the square
of the exotic amplitude, which should be strongly suppressed with
respect to the previous one. An approximate formula for $|\sigma_{
\kappa^2}|$, valid only at the first non-trivial order in the
expansion parameter $(2 \, m_W / \sqrt{s})$, but sufficient for an
order of magnitude estimate, is:
\be
\label{swwapp}
\left| \sigma_{\kappa^2} \right| \simeq {\alpha \over 768 
\, \sin^2 \theta_W \, \cos^2 \theta_W} \ {s \over M^4} \, .
\ee
Taking $M$ at its lower bound (\ref{zinvbound}), and $\sqrt{s} = 206.6
\ {\rm GeV}$, where the LEP2 average \cite{lepewwg} is
$\sigma_{WW}^{(LEP)} = 17.28 \pm 0.27 \ {\rm pb}$, from (\ref{swwapp})
we find $\sigma_{\kappa^2} \simeq 0.18 \ {\rm pb}$, well below one
standard deviation. We have checked that the complete theoretical
expression for $|\sigma_{\kappa^2} + \sigma_{\kappa^4}|$ gives a
similar constraint, and that data at lower values of $\sqrt{s}$ are
less restrictive.

It is conceivable that the new physics originating the operator
(\ref{operator}) gives rise to additional operators in the effective
theory at the weak scale, for example local operators involving only
four SM fermions. In particular, four-fermion operators involving at
least two left-handed leptons of type $\widehat{a}$ can be generated
by quantum corrections (e.g. box diagrams) in the presence of the
interactions (\ref{due}) and (\ref{tre}). In the generic framework of
models with a low supersymmetry-breaking scale, the question of
four-SM-fermion operators was addressed in \cite{bfz4f}.  It was found
that a supersymmetry-breaking scale as low as the direct experimental
bound can naturally coexist with the level of suppression of the
dangerous four-fermion operators required by the LEP and Tevatron
bounds \cite{pdg}.

A similar question could be asked in the more specific framework of
superstring models with a low string scale \cite{strsca}. The
strongest bounds arise in the case where all four fermions are
localized at the same brane intersection, giving rise to dimension six
effective operators. However, deciding whether those bounds could be
applicable in the present context would require an explicit realistic
string construction. To be conservative, we should consider only
four-fermion operators involving two left-handed leptons of type
$\widehat{a}$. The strength of the corresponding dimension-six
effective operator depends on the intersection angle, and becomes
maximal for orthogonal branes. In the case of coincident branes, the
dimension-six effective operator vanishes, and the leading
contribution to four-fermion contact interactions comes at dimension
eight. The limit on the string scale can then be extracted from Bhabha
scattering, leading to $M_s > 490$ GeV \cite{strlep}. This is roughly
comparable with the direct bound (\ref{zinvbound}), taking into
account that, as argued in section~1.2, we expect $M_s \sim 1.6 \
\sqrt{F} \sim 1.9 \ M$.

To conclude, we notice that the new interactions in (\ref{due}) and
(\ref{tre}) do not contribute to Veltman's $\rho$ parameter at the
one-loop level. There are, however, quadratically divergent one-loop
contribution to $[\Pi_{VV}^{new} (m_V^2) - \Pi_{VV}^{new} (0) ] /
m_V^2$, ($V=Z,W$), originated by gauge boson self-energy diagrams with
a goldstino and a lepton of type $\widehat{a}$ on the internal lines.
Choosing an ultraviolet cutoff $\Lambda \sim M$ suggests that the
natural value of these contributions is ${\cal O} [ m_V^2 / (64 \,
\pi^2 \, M^2)]$. Even for $M$ at its lower bound (\ref{zinvbound}) and
$V=Z$, we find an ${\cal O} (1.8 \times 10^{-4})$ contribution,
corresponding to $\delta \, S = \cO ( 1.6 \times 10^{-2} )$ or
$\delta \, \widehat{\epsilon}_3 = \cO (1.4 \times 10^{-4} )$, in
agreement with the present bounds \cite{pdg}.

\subsection{Higgs boson decays}

The constraints discussed in the previous section leave room for very
interesting signatures of the new goldstino interactions in Higgs
decays.

The interaction in (\ref{uno}) describes the invisible decays $H
\rightarrow \nu \, \widetilde{G}$ and $H \rightarrow \ov{\nu} \,
\ov{\widetilde{G}}$, at a rate:
\be
\Gamma_{inv} (H) =
\Gamma ( H \rightarrow \nu \, \widetilde{G} ) + 
\Gamma ( H \rightarrow \ov{\nu} \, \ov{\widetilde{G}} ) =
{m_H^5 \over 64 \ \pi \ M^4} \, .
\label{hdecay}
\ee
Notice the similarity between eq.~(\ref{hdecay}) and the universal
formula \cite{cfm} expressing the decay rate for a massive
superparticle into its massless superpartner and a goldstino. Indeed,
using the spinor algebra, integration by parts and the equations of
motion, the three-point coupling (\ref{uno}) can be rewritten as
$[1/(2 \, \sqrt{2})] \, \kappa \, c \, m_H^2 \, \widetilde{G} \, \nu
\, H + {\rm h.c.}$, as if the spin-0 Higgs doublet and the spin-$1/2$
lepton doublet were members of a single chiral supermultiplet in the
linear realization. It is a curious coincidence that, in the D-brane
models of \cite{anttuc}, Higgs and lepton doublets sit in the same
multiplet, but of a supersymmetry different from the one associated
with the goldstino $\widetilde{G}$.
 
The interactions in (\ref{quattro}) and (\ref{cinque}) could describe
the decays $H \rightarrow Z \, \nu \, \widetilde{G}$, $Z \, \ov{\nu}
\, \ov{\widetilde{G}}$, $W^- \, l^+ \, \ov{\widetilde{G}}$, $W^+ \,
l^- \, \widetilde{G}$, which however are strongly suppressed by phase
space with respect to the two-body decays in (\ref{hdecay}), and will
be neglected here.

The present LEP2 bound on a Higgs boson with SM production
cross-section but dominant invisible decays \cite{invhbound} is very
close to the bound on the SM Higgs \cite{smhbound}, $m_H > 114.4 \
{\rm GeV}$ at $95 \, \% \ {\rm c.l.}$. In the following we will then
consider values of the Higgs boson mass above the SM bound, even if
the precise bound on $m_H$ for a Higgs boson with both SM and
invisible decay modes could be slightly weaker. We also remind the
reader that, in models with a low supersymmetry breaking scale, and in
contrast with the MSSM, a SM-like Higgs boson is not bound to be light
(for a recent discussion, see e.g. \cite{bcen}).

Taking into account the constraint (\ref{zinvbound}) and the
experimental lower bound on $m_H$, we can now study the possible
phenomenological relevance of the new invisible Higgs decay modes of
eq.~(\ref{hdecay}). We computed the Higgs branching ratios, as
functions of $m_H$ and $M$, by including the new invisible channels in
the program HDECAY \cite{HDECAY}. The results are displayed in
figs.~1--4. Fig.~1 shows the most important Higgs branching ratios as
functions of $m_H$, for two representative values of $M$: 300 and
600~GeV. Fig.~2 shows the same branching ratios, but as functions of
$M$, for four representative values of $m_H$: 115, 140, 200 and
400~GeV. Fig.~3a shows contours of $BR_{inv} (H) = \Gamma_{inv} (H) /
\Gamma_{tot} (H)$ in the $(M,m_H)$ plane. Fig.~3b shows the ratio
$\Gamma_H / m_H$, as a function of $m_H$, in the large mass region, in
the SM case and for two representative values of $M$: 300 and 500~GeV
(notice that the 500~GeV line is already very close to the SM one).
Figs.~4a and 4b show the total Higgs width $\Gamma_H$ and the
branching ratio $BR ( H \to \gamma \, \gamma )$ as functions of $m_H$,
in the SM case and for the same representative values of $M$ as in
Fig.~3b.

We can see from the various figures that, for values of $M$ close to
the lower bound of eq.~(\ref{zinvbound}), the invisible Higgs decay
modes can be the dominant ones. This behaviour persists for moderate
values of $M$, especially for $m_H \sim 130$~GeV, where we can still
have an ${\cal O}(10 \%)$ invisible branching ratio for $M \sim 750 \
{\rm GeV}$. When the invisible decay modes dominate, we could see an
effect both in a dedicated search for invisible Higgs decays, and,
indirectly, by measuring a deficit in the Higgs branching ratios into
SM channels, or a total Higgs width larger than the SM prediction.

We should recall at this point that in models where $\sqrt{F}$ is
close to the weak scale but supersymmetry is linearly realized, the
goldstino can pick up small components along the neutralinos (gauginos
and higgsinos). These can in turn induce $Z$ \cite{lutpon, bcen} and
Higgs \cite{bcen} couplings to goldstino pairs, and lead to the decays
$Z \to \widetilde{G} \, \ov{\widetilde{G}}$ or $H \to \widetilde{G} \,
\widetilde{G}$ at a non-negligible rate. Here, however, we work in the
non-linear realization, and we assumed that the goldstino is a pure SM
singlet, thus we consistently neglected such a possibility.  The
non-linear realization also permits, in principle, a dimension-seven
gauge-invariant operator \cite{anttuc} proportional to $[ \phi^\dagger
\phi \ (\partial_\mu \widetilde{G}) \sigma^{\mu \nu} (\partial_\nu
\widetilde{G}) + {\rm h.c.}]$, which would violate total lepton number
and lead to $H \to \widetilde{G} \, \widetilde{G}$ decays. Here,
according to the results of explicit calculations in D-brane models
\cite{anttuc}, we assumed that the above dimension-seven operator is
absent.

The possibility of invisible Higgs decays was also considered in other
theoretical frameworks. One is the existence of a fourth generation,
with negligible mixing with the first three, and a specific mass
spectrum: this may allow for $H \to N \, \ov{N}$ decays \cite{khl},
where $N$ is a 50-80~GeV neutrino, but is only marginally allowed by
electroweak precision data \cite{norv, pdg}. Another one is the MSSM
with non-universal gaugino masses, which may allow for a large $H \to
\widetilde{\chi}^0 \, \widetilde{\chi}^0$ branching ratio, where
$\widetilde{\chi}^0$ is the lightest neutralino. A third one is the
possibility of Higgs-radion mixing in models with extra dimensions
\cite{radhig}: in that case, however, the invisible channel can never
be the dominant decay mode. A final one is the possibility of Higgs
decays into Majorons in non-minimal supersymmetric models with
spontaneously broken $R$-parity \cite{major}: in this case, at the
price of rather complicated constructions, the invisible mode could be
dominant.

\section{Conclusions and outlook}

In this paper we examined the phenomenological implications of the
dimension-six operator (\ref{operator}), allowed by non-linear
supersymmetry coupled to the minimal non-supersymmetric SM. We worked
in the limit where the new couplings involve only one lepton family
and neutrino masses are vanishing, so that total and partial lepton
numbers are conserved in perturbation theory. We showed that the most
stringent phenomenological constraints on the mass scale $M$ of the
new interactions come from invisible $Z$ decays, and give $M > 270 {\;
\rm GeV}$. We also examined other constraints from measured processes,
and found that either they are less stringent or they have a more
ambiguous physical interpretation. We finally discussed the most
striking phenomenological signature originated by the new
interactions: the possibility for the Higgs boson to decay into the
invisible channel neutrino $+$ goldstino, or the conjugate one, with
non-negligible or even dominant branching ratio.

The phenomenological scenario discussed in this paper can be tested at
the Tevatron, at the LHC, and at a possible international $e^+ e^-$
linear collider (ILC) with $\sqrt{s} \simgt 500 {\; \rm GeV}$
\cite{ilc}.  These colliders can improve the existing bounds on
anomalous interactions among SM fermions and on anomalous
contributions to $W$-pair production, albeit with the interpretation
ambiguities already mentioned in subsection~2.1. More importantly, Run
II of the Tevatron and the LHC can probe higher values of the
supersymmetry breaking scale $\sqrt{F}$, by looking for single photon
(single jet) plus missing transverse energy, as originated by the
production of goldstino-antigoldstino pairs in association with a
photon (jet). However, extracting reliable experimental bounds on
$\sqrt{F}$ at hadron colliders seems to be more difficult than
expected by preliminary theoretical studies \cite{tevth}. The latter
estimated the Tevatron Run~I sensitivity at $\sqrt{F} \sim 310 \ {\rm
GeV}$, whereas the actual experimental bound \cite{tevbound} is only
$\sqrt{F} > 221 {\; \rm GeV}$. Preliminary theoretical estimates
\cite{tevth} for Run II of the Tevatron and for the LHC were
foreseeing a sensitivity up to $\sqrt{F} \sim 410 \ {\rm GeV}$ and
$\sqrt{F} \sim 1.6 \ {\rm TeV}$, respectively, but taking into account
the actual detector environment and the very similar shapes of signal
and background may lead to a significant downgrade of these
estimates. The two main processes of interest for the ILC are:
goldstino pair production in association with a photon, which should
be sensitive to values of $\sqrt{F}$ of the order of $\sqrt{s}$; the
production of neutrino-goldstino pairs in association with a photon,
for which one could repeat the LEP2 analysis of
eqs.~(\ref{dsph})--(\ref{sigma0th}), eventually dropping the
simplifying approximations.

As for invisible Higgs decays, looking for direct or indirect signals
at the LHC looks quite challenging but not impossible. For example,
having dominant decays into invisible channels does not seem to
preclude the possibility of Higgs detection, even if it makes it more
difficult: existing proposals try to exploit associated $ZH$
production \cite{roy}, associated $t \, \ov{t} \, H$ production
\cite{tthinv}, production via $WW$ fusion with tagged forward jets
\cite{wwhinv}, and central exclusive diffractive production
\cite{difhinv}. Indirect signals would correspond to measuring
deviations from the SM predictions for the Higgs branching ratios into
SM channels and for the total Higgs width. Direct and indirect
detection of an invisible Higgs decay channel should be much simpler
at an ILC with sufficient energy and luminosity.

There are many generalizations of the present analysis that would be
interesting to consider. A more systematic study of the processes
considered in section~2.1 could be performed. The possibility of
generating the operator $\cO$ starting from an underlying linear
realization could be also explored: one may be led to a two-Higgs
model with some operators breaking lepton number and $R$-symmetry to
some diagonal subgroup; alternatively, taking into account the hints
coming from brane-world models, one may consider a model with fully
broken extended supersymmetry and additional massive states besides
those of the MSSM.

Another important generalization, in analogy with the studies of
$R$-parity breaking in the MSSM, would be the discussion of the
flavour-dependent phenomenology in the presence of more than one
non-negligible $c_a$ ($a=e, \, \mu, \, \tau$). This could be also
related with attempts at addressing the flavour problem within
semirealistic brane models \cite{bramod}. In such a case, the operator
$\cO$ still conserves the total lepton number but violates lepton
flavour.  However, lepton flavour violation (LFV) appears in the very
special form dictated by the operator (\ref{operator}), which would
deserve a dedicated analysis of the processes with LFV. The discussion
of neutrino masses and neutrino oscillations would presumably be
non-trivial. Introducing Majorana masses for the left-handed neutrinos
via dimension-five operators, we would have no new dimension-six
operators besides (\ref{operator}), but we would need to consider both
the latter and neutrino masses as sources of LFV. Introducing
right-handed neutrinos into the model, to generate Dirac neutrino
masses via Yukawa couplings, would force us to consider an additional
dimension-six operator besides (\ref{operator}), of the form
\be
\label{optilde}
\widetilde{\cO} = i \ \kappa \ \sum_{a=e,\mu,\tau} \widetilde{c}_a 
\ \widetilde{\cO}_a + h.c. \, ,
\qquad
\widetilde{\cO}_a =  \nu^c_a \ \sigma^\mu \ (\partial^\nu 
\ov{\widetilde{G}}) \ B_{\mu \nu} \, ,
\ee
where $B_{\mu \nu}$ is the $U(1)_Y$ field strength and string
considerations suggest this time $\widetilde{c}_a = 0$ or $|
\widetilde{c}_a | = \sqrt{2}$. All this, however, goes beyond 
the aim of the present paper and is left for future work.

\newpage

\section*{Acknowledgments}

We would like to thank J.-P.~Derendinger, F.~Feruglio, P.~Gambino,
M.~Khlopov, C.~Kounnas, S.~Lammel, M.~Mangano, R.~Rattazzi,
A.~Romanino, D.P.~Roy, A.~Strumia for discussions, and A.~Brignole for
discussions and valuable comments on a preliminary version of this
paper. This work was supported in part by the European Commission
under the RTN contracts HPRN-CT-2000-00148 (Across the Energy
Frontier) and MRTN-CT-2004-503369 (Quest for Unification).

\newpage
\section*{Figures}
\vspace*{2.0cm}
\begin{figure}[ch]
\begin{center}
\mbox{
\hspace{-0cm}
\epsfig{figure=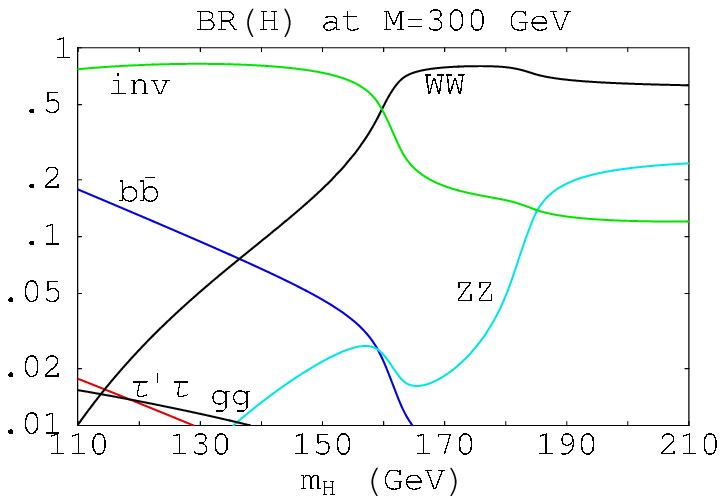,width=7.5cm,height=7.5cm}
\hspace{-0cm}
\epsfig{figure=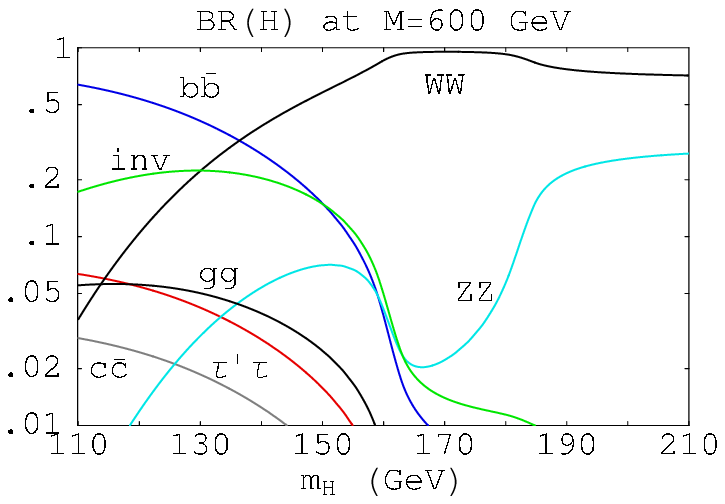,width=7.5cm,height=7.5cm}
}
\end{center}
\vspace{0cm}
\begin{center}
\mbox{
\hspace{-0cm}
\epsfig{figure=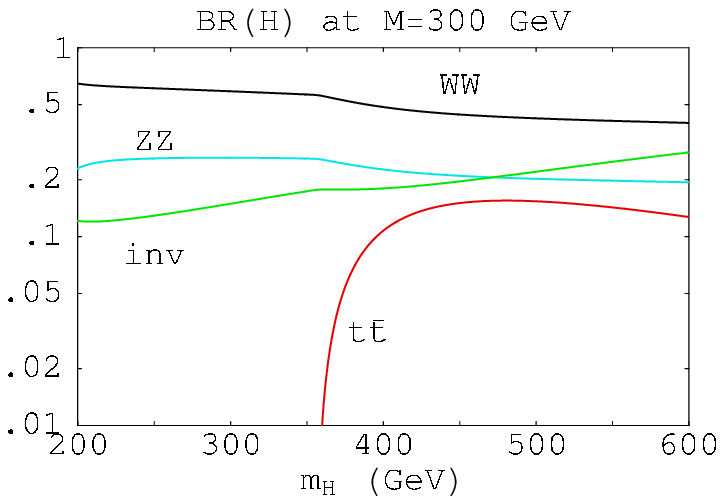,width=7.5cm,height=7.5cm}
\hspace{-0cm}
\epsfig{figure=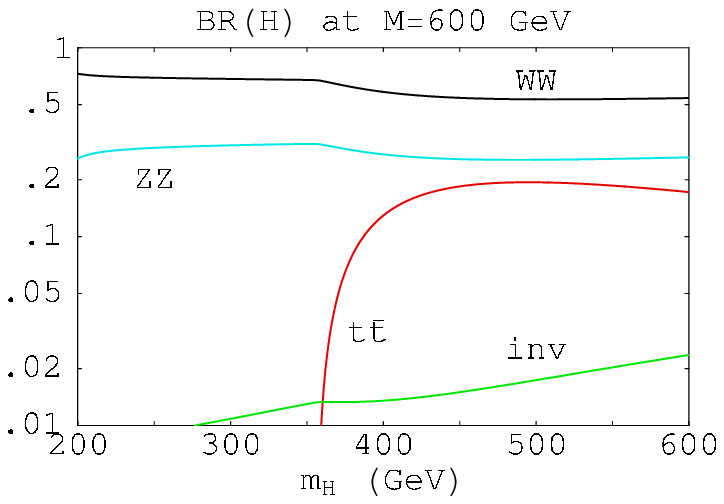,width=7.5cm,height=7.5cm}
}
\end{center}
\vspace{-0.6cm}
\caption{The most important Higgs branching ratios, as functions of
$m_H$, for two representative values of $M$: 300 and 600~GeV.}
\label{fig1}
\end{figure}
\newpage
\begin{figure}[c]
\begin{center}
\mbox{
\hspace{-0cm}
\epsfig{figure=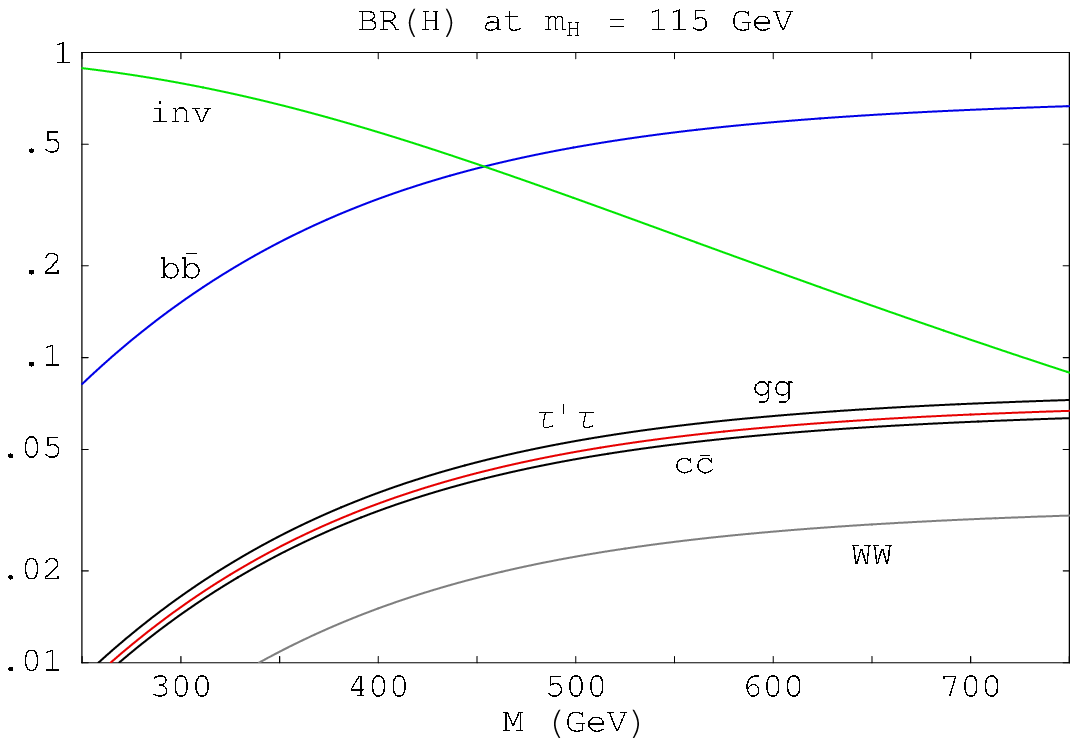,width=7.5cm,height=7.5cm}
\hspace{-0cm}
\epsfig{figure=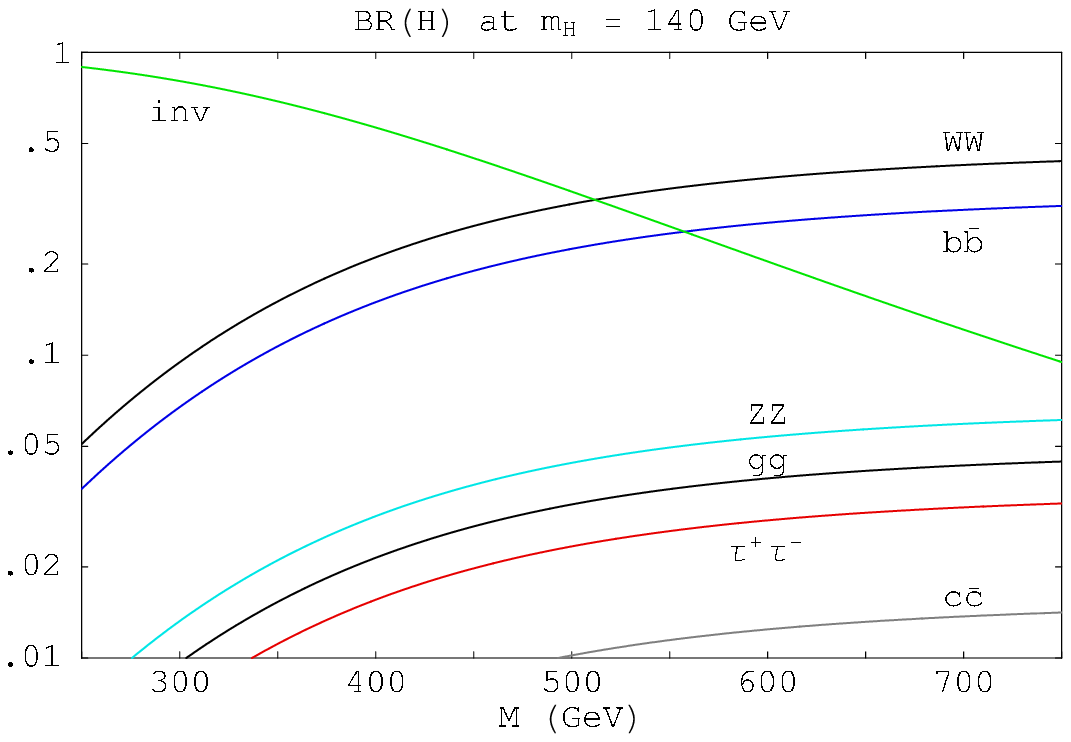,width=7.5cm,height=7.5cm}
}
\end{center}
\vspace{-0cm}
\begin{center}
\mbox{
\hspace{-0cm}
\epsfig{figure=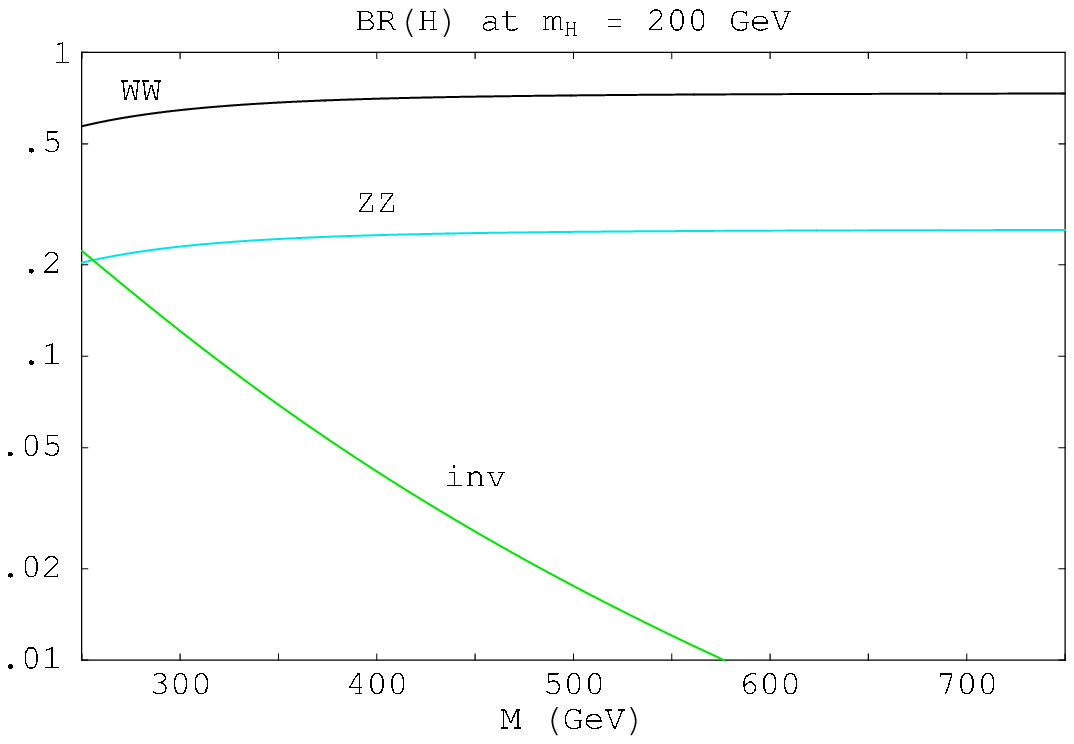,width=7.5cm,height=7.5cm}
\hspace{-0cm}
\epsfig{figure=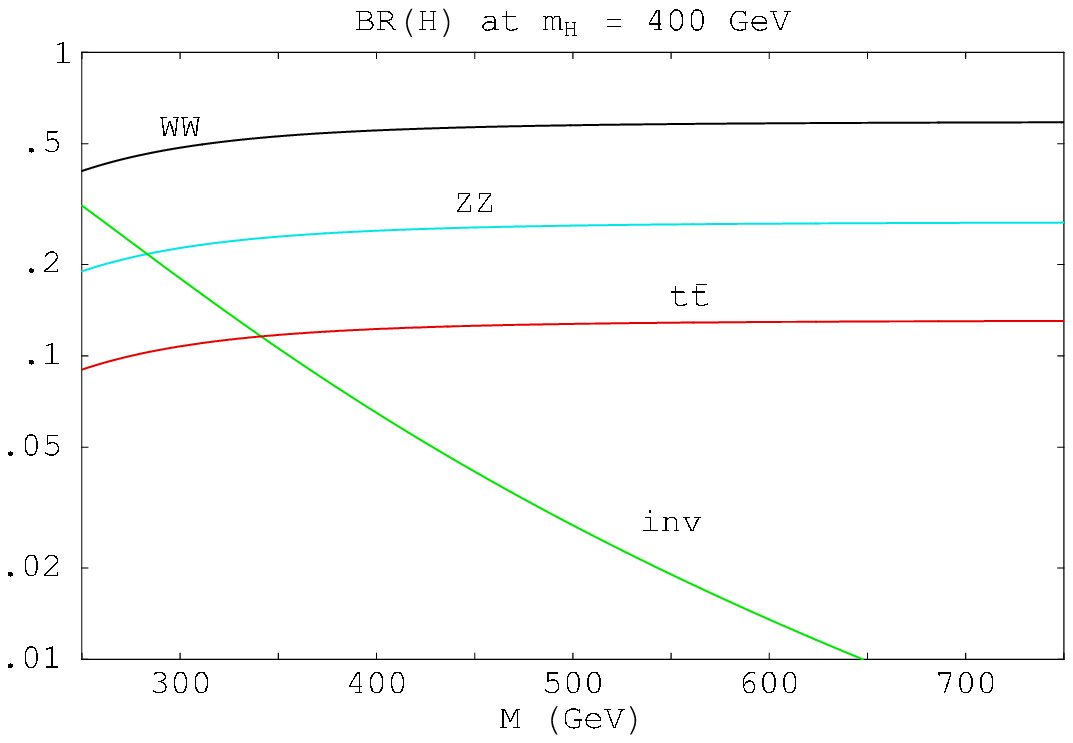,width=7.5cm,height=7.5cm}
}
\end{center}
\vspace{-0.6cm}
\caption{The most important Higgs branching ratios, as functions of
$M$, for four representative values of $m_H$: 115, 140, 200 and
400~GeV.}
\label{fig2}
\end{figure}
\newpage
\begin{figure}[htb]
\begin{center}
\mbox{
\hspace{-0.4cm}
\epsfig{figure=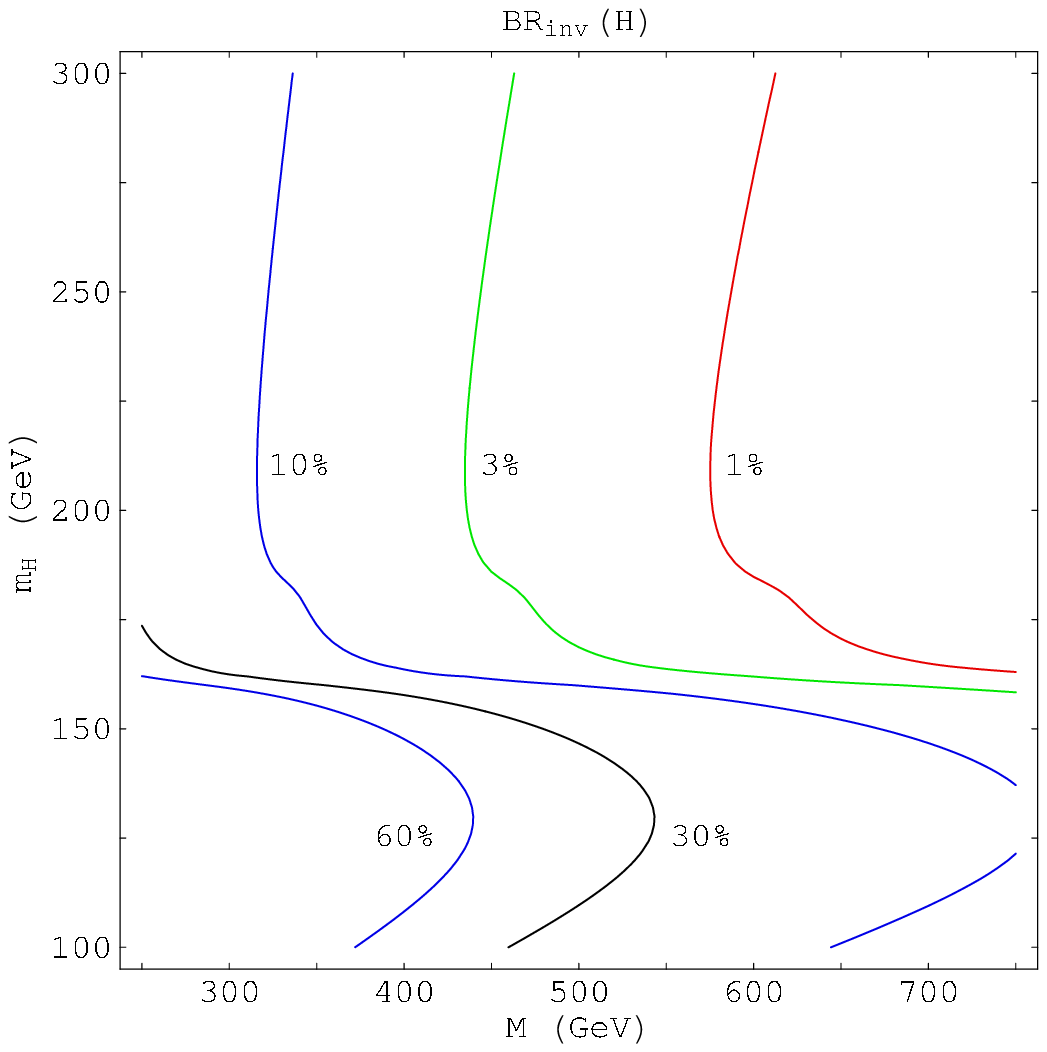,width=7.5cm,height=7.5cm}
\hspace{-0cm}
\epsfig{figure=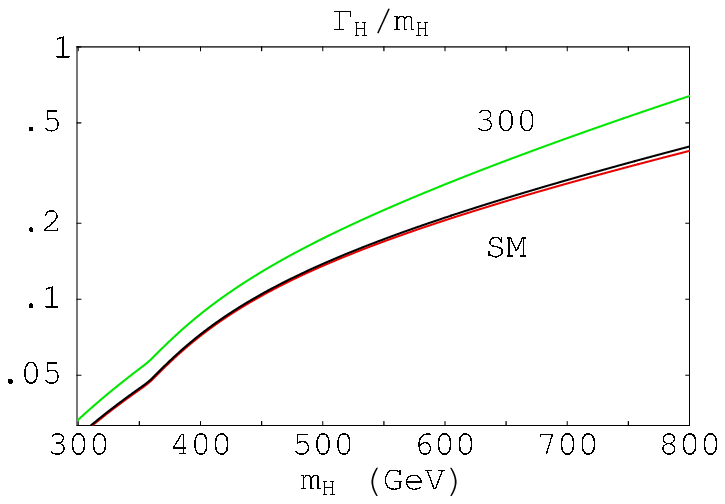,width=7.5cm,height=7.5cm}
}
\end{center}
\vspace{-0.6cm}
\caption{(a) Contours of the invisible Higgs branching ratio $BR_{inv}
(H)$ in the $(M,m_H)$ plane. (b) The ratio $\Gamma_H / m_H$, as a
function of $m_H$ in the large mass region, in the SM and for two
representative values of $M$: 300 and 500~GeV.}
\label{fig3ab}
\end{figure}
\begin{figure}[hbt]
\begin{center}
\mbox{
\hspace{-0.4cm}
\epsfig{figure=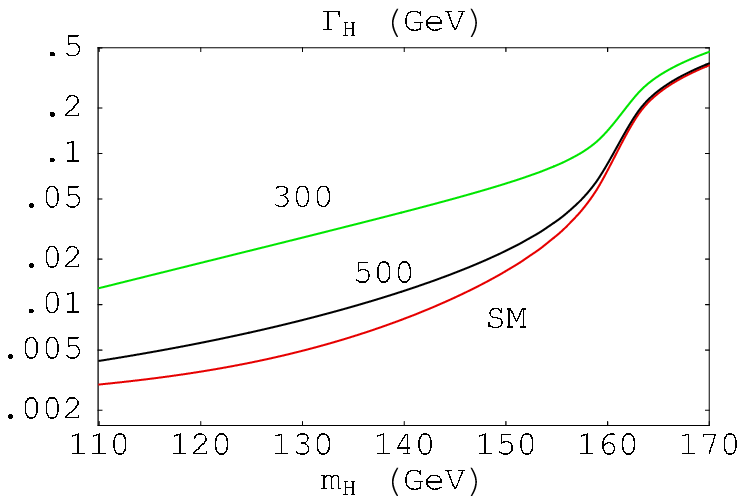,width=7.5cm,height=7.5cm}
\hspace{-0cm}
\epsfig{figure=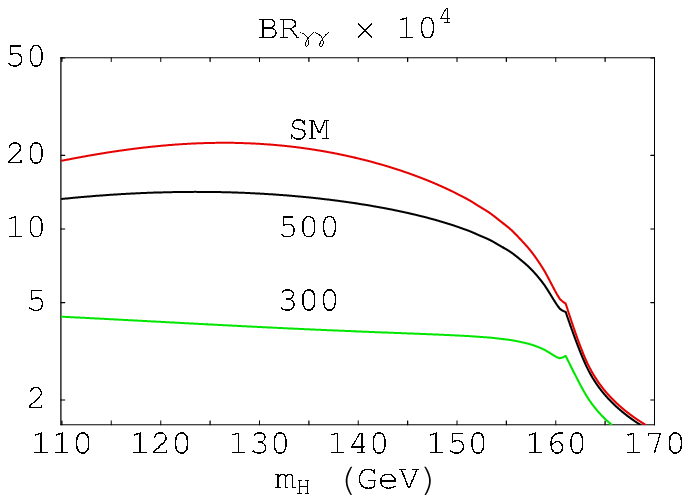,width=7.5cm,height=7.5cm}
}
\end{center}
\vspace{-0.6cm}
\caption{The total Higgs width $\Gamma_H$ and the branching 
ratio $BR(H \to \gamma \, \gamma)$, as functions of $m_H$
in the intermediate mass region, in the SM and for two 
representative values of $M$: 300 and 500~GeV.}
\label{fig4ab}
\end{figure}
\end{document}